\documentclass[%
 reprint,
 onecolumn,
superscriptaddress,
 amsmath,amssymb,
 aps,
nofootinbib,floats
]{revtex4-2}

\usepackage[letterpaper,top=2cm,bottom=2cm,left=3cm,right=3cm,marginparwidth=1.75cm]{geometry}

\usepackage{graphicx}% Include figure files
\usepackage{dcolumn}% Align table columns on decimal point
\usepackage{bm}% bold math
\usepackage[hidelinks]{hyperref}

\usepackage{xcolor}

\usepackage{soul}
\usepackage[normalem]{ulem}

\begin{document}

\title{Interplay of viscosity and wettability controls fluid displacement in porous media}

%%% RH: article
% \author[1]{Saideep Pavuluri}
% \author[2]{Ran Holtzman}
% \author[1]{Luqman Kazeem}
% \author[1]{Malyah Mohammed}
% \author[1]{Thomas Daniel Seers}
% \author[1]{Harris Sajjad Rabbani}
% \affil[1]{Department of Petroleum Engineering, Texas A\&M University at Qatar, Education City, Doha, Qatar}
% \affil[2]{Fluid and Complex Systems Research Centre, Coventry University, Coventry CV1 2NL, United Kingdom}
%%% RH: APS
\author{Saideep Pavuluri}
%\email{saideep025@gmail.com}
\affiliation{Department of Petroleum Engineering, Texas A\&M University at Qatar, Education City, Doha, Qatar}
\author{Ran Holtzman}
%\email{ran.holtzman@coventry.ac.uk}
\affiliation{Fluid and Complex Systems Research Centre, Coventry University, Coventry, United Kingdom}
\author{Luqman Kazeem}
\affiliation{Department of Petroleum Engineering, Texas A\&M University at Qatar, Education City, Doha, Qatar}
%\email{}
\author{Malyah Mohammed}
\affiliation{Department of Petroleum Engineering, Texas A\&M University at Qatar, Education City, Doha, Qatar}
%\email{}
\author{Thomas Daniel Seers}
\affiliation{Department of Petroleum Engineering, Texas A\&M University at Qatar, Education City, Doha, Qatar}
%\email{}
\author{Harris Sajjad Rabbani}
\email{harris.rabbani@qatar.tamu.edu}
\affiliation{Department of Petroleum Engineering, Texas A\&M University at Qatar, Education City, Doha, Qatar}

%%% RH: article STYLE
% \begin{document}
% RH \doublespacing
% RH \maketitle
% replace \sout by \st \usepackage{soul} vs \usepackage[normalem]{ulem}

%\newpage

\begin{abstract}
Direct numerical simulations are used to elucidate the interplay of wettability and fluid viscosities on immiscible fluid displacements in a heterogeneous porous medium.
%during viscous-dominated flow conditions.
%
% We simulate the displacement of fluids of a wide range of viscosity ratios ($M$, ratio of invading to defending fluid viscosity) and wettability 
% %ranging from highly wetting to nonwetting invading fluid 
% in a geologically-realistic two-dimensional pore space obtained from X-ray tomography image of a sand pack. 
%
We classify the flow regimes based 
%on the simulated displacement patterns, 
using qualitative and quantitative analysis into viscous fingering (low $M$), compact displacement (high $M$), and an intermediate transition regime ($M \approx 1$). 
We use stability analysis to obtain theoretical phase boundaries between these regimes, which agree well with our analyses.
At the macroscopic (sample) scale, we find that wettability strongly controls the threshold $M$ (at which the regimes change).
%, which increases with the contact angle.
%between the fluid-fluid interface and solid surface. 
At the pore scale, wettability alters the dominant pore-filling mechanism. At very small $M$ (viscous fingering regime), smaller pore spaces are preferentially invaded during imbibition, with flow of films of invading fluid along the pore walls. 
In contrast, during drainage, bursts result in filling of pores irrespective of their size. 
As $M$ increases, the effect of wettability decreases as cooperative filling becomes the dominant mechanism regardless of wettability. 
This suggest that for imbibition at a given contact angle, decreasing $M$ is associated with change in effective wetting from neutral-wet (cooperative filling) to strong-wet (film flow).
\end{abstract}

\maketitle


%APS  does not ask for this so I comment for now; https://journals.aps.org/prfluids/authors
% \begin{figure}[!]
%     \centering
%     \includegraphics*[scale=0.25]{Figures/abstractFig.pdf}
%  %   \caption{Graphical abstract (if required / allowed).}
%  %   \label{fig:my_label}
% \end{figure}
%\newpage

%-------------------%
\section{Introduction}
\label{Intro}

Fundamental understanding of immiscible fluid-fluid displacements in porous media is vital for the safe and efficient operation of a large number of engineering applications. Examples include sequestration of carbon dioxide \citep{metz2005ipcc, boot2014carbon}, the fate of non-aqueous phase liquid contaminants (NAPLs) in groundwater \citep{villaume1985investigations, mayer2005soil} and their remediation \citep{mackay1989groundwater}, geothermal energy \citep{o2001state, barbier2002geothermal} and extraction of hydrocarbons \citep{sheng2010modern,cyr2001steam}. 
The displacement patterns are controlled by the interplay of capillary, viscous, inertial and gravitational forces, which in turn are controlled by a wide range of parameters, including flow rates, fluid properties (viscosity, density, surface tension), wettability, and medium properties (pore sizes, shapes and connectivity) \citep{avraam1995flow}.

The classical phase diagram of fluid displacement regimes predicts the transition between compact displacements, viscous fingering and capillary fingering patterns considering the capillary number $Ca$ (ratio of viscous to capillary forces) and viscosity ratio $M$ \citep{lenormand1988numerical, lenormand1990liquids}. 
%\sout{, for either perfectly wetting or perfectly non-wetting fluids.}
%
% \textcolor{brown}{
% In general, for viscous fingering and compact displacements the viscous forces play a relatively significant role in controlling the flow dynamics, whereas, for capillary fingering, the capillary forces play a critical role in controlling the morphology of fluid displacements.}
% %
%\textcolor{brown}{
These seminal works did not consider however the effect of wettability---the affinity of one fluid relative to another to adhere to the solid surface, measured via the contact angle $\theta$. Instead, they considered only the extreme cases of drainage and imibibition for perfect wetting, that is the displacement of a perfectly nonwetting and wetting fluid, respectively. 
Recent works however expose the crucial impact of wettability, showing it substantially affects the pore filling mechanisms and thus the displacement patterns \citep{trojer2015stabilizing, holtzman2015wettability, zhao2016wettability, rabbani2017new, rabbani2018pore, hu2018wettability, pavuluri2020towards, zakirov2021wettability}.

To fill this gap, a few studies used numerical simulations to include the effect of wettability on displacement morphology. %and fluid viscosities. 
% I would put primkuluv first as the most comprehensive, with Ca
\citet{primkulov2021wettability} used dynamic pore network modelling to decipher the interplay between wettability, viscosities and flow rates, extending the classical phase diagram in \cite{lenormand1988numerical}.
%to include the contact angle. 
The authors considered an idealized porous medium made of cylindrical pillars, for which analytical expressions providing the advancement of the meniscus via various filling mechanisms can be established \cite{primkulov2018quasistatic}.
Lattice Boltzmann (LB) simulations were used to study the impact of wettability and viscosity on the displacement efficiency for viscous fingering, on a similar idealized medium \citep{mora2021influence}. 
%A similar medium was considered in a comprehensive study that exposed the interplay between wettability, viscosities and flow rates (capillary number), extending the classical phase diagram in \citet{lenormand1988numerical} to include the contact angle \citep{primkulov2021wettability}. The authors used pore network modeling, exploiting analytical expressions providing the advancement of the meniscus via various filling mechanisms, which exist for the special case of solid particles shaped as cylindrical pillars \citep{primkulov2018quasistatic}.
%
%%% RH \cite{primkulov2021wettability} used dynamic pore network modelling to decipher the interplay between wettability, viscosities and flow rates (capillary number), extending the classical phase diagram in \cite{lenormand1988numerical} to include the contact angle. The authors considered an idealized porous medium made of cylindrical pillars, for which analytical expressions providing the advancement of the meniscus via various filling mechanisms can be established \cite{primkulov2018quasistatic}.
%
A more realistic 3-D pore geometry, extracted from from micro-CT images of sandstone, was used in LB simulations to study how wettability and geometrical pore-scale heterogeneity affects the displacement, for two different viscosity ratios, favorable and unfavorable ($M>>1$ and $M<<1$, respectively) \citep{bakhshian2021physics}. 
These works showed a transition from viscous fingering (VF) to compact displacement (CD) with increasing $M$, and the existence of an intermediate (VF/CD) regime, akin to the transition regime between CF and CD found earlier at lower flow rates (lower $Ca$) by \citet{hu2018wettability}.

Here, we systematically explore the synergistic relationships between wettability and fluid viscosities in a geologically realistic medium under viscous-dominated flow. We use Direct Numerical Simulations (DNS), which allow consideration of the physics (solving the fundamental flow equations) for the intricate pore geometry derived from a micro-CT image of a sand pack. The high spatial and temporal resolution provided by DNS can capture sub-pore scale events such as interfacial readjustments \citep{ferrari2014inertial, pavuluri2020towards}, cooperative filling \citep{pavuluri2019direct}, flow of films along the solid surfaces \citep{abu2017multiscale}, and non monotonic behaviour of the capillary pressure \citep{rabbani2016effects, rabbani2019inertia, pavuluri2019direct}. This allows us to relate the large-scale features of invasion patterns and displacement efficiency to the pore-scale mechanisms controlling it. 
%To be able to run multiple realizations of sufficiently large domains, we simulate flow in 2-D cross-sections in light of the large computational costs of DNS. 
Our simulations capture the different displacement regimes (VF, CD and VF/CD) 
%, namely, compact displacements, intermediate regime, and viscous fingering. We found 
showing that the threshold $M$ for the crossover between the regimes increases with the wettability. 
We also establish the dependence of the dominant pore filling mechanisms on the combination of $M$ and wettability. In particular, we show that for imbibition at a given contact angle, the effective wettability changes with viscosity: from neutral-wetting (cooperative filling) to strongly-wetting (film flow) with decreasing $M$.

\section{Mathematical Model}
\label{Mathmodel}
%The set of mathematical equations that are solved numerically to investigate multiphase flows in the pore spaces are initially discussed. We then briefly describe the heterogeneous porous medium that has been used in this work to investigate immiscible fluid displacement patterns. Later, we discuss the boundary conditions used for the numerical set-up followed by description of the model parameters.

\subsection{Governing equations}
%\subsection{Physical and Numerical Description of Multiphase Flows at the Pore-Scale}
\label{DNSEqs}
In DNS, the isothermal flow dynamics of immiscible and incompressible multiphase flow systems are governed by the Navier-Stokes equations, solved for each fluid phase, where the fluid-fluid interfaces boundary conditions that ensure continuity in the velocity field are set. The stress gradients can be computed via the Young-Laplace equation \citep{batchelor2000introduction}. Though this approach provides the interface dynamics, it is computationally demanding as it requires solving for a complex moving boundary problem. 
The Volume of Fluid (VOF) method simplifies the computations by considering the two fluid phases as a single mixture \citep{hirt1981volume}. This is done by defining a colour function $\alpha \in [0, 1]$ which indicates the volume occupied by a specific fluid in a control volume: when $\alpha$ is equal to 0 or 1 the volume is occupied by a single phase, whereas $1 > \alpha >0$ indicates the co-existence of two fluids seperated by at least one fluid-fluid interface. 
VOF then solves for conservation of both mass
\begin{equation}
    \label{eq:massBal}
    \nabla \cdot \textbf{U} = 0
\end{equation}
and momentum
\begin{equation}
    \label{eq:momBal}
    \frac{\partial (\rho \textbf{U})}{\partial t} + \nabla \cdot (\rho \textbf{UU}) = -\nabla p + \nabla \cdot \mu (\nabla \textbf{U} + \nabla \textbf{U}^{T}) + \textbf{F}_{bdy} + \textbf{F}_{cap}.
\end{equation}
In Eqs. \ref{eq:massBal}--\ref{eq:momBal}, $\textbf{U}$ is the velocity, $t$ is time, $p$ is pressure, $\textbf{F}_{bdy}$ and $\textbf{F}_{cap}$ are the external body (for example: gravity) and capillary forces, respectively. Superscript $T$ denotes a transpose.
%
%%% In VOF, $\textbf{U} = \textbf{U}_{n} = \textbf{U}_{w}$ where the subscripts $n, w$ refer to the non-wetting and wetting phases respectively. 
%[RH: unclear, explain! is this on the interface?]
Considering a single mixture, its properties are assumed to be a linear combination of the two fluids comprising it; for example, denoting the wetting phase by $\alpha = 1$, the density $\rho$ and the dynamic viscosity $\mu$ of the mixture are
\begin{equation}
\label{eq:rho_mu}
    \begin{aligned}
        \rho = \rho_{w} \alpha + \rho_{n} (1 - \alpha), \\
        \mu = \mu_{w} \alpha + \mu_{n} (1 - \alpha).
    \end{aligned}
\end{equation}

The capillary forces are 
\begin{equation}
\label{eq:Fcapbdy}
    \textbf{F}_{cap} = \sigma k \textbf{n}_{I} \delta_{I}
\end{equation}
where $\sigma$ is the surface tension, $k$ is the interface curvature, $\textbf{n}_{I}$ is the unit normal to the interface and $\delta_{I}$ is a Dirac delta function, which is used to restrict the capillary forces to act only at the interface. 
Various VOF formulations can be used to define how the normal to the interface $\textbf{n}_{I}$ and the Dirac delta function $\delta_{I}$ are discretized, see \citet{pavuluri2018spontaneous}. In this work we use the conventional Continuum Surface Force (CSF) formulation \citep{brackbill1992continuum}, in which $\textbf{n}_{I}\delta_{I}$ from Eq. \eqref{eq:Fcapbdy} 
%in Eq. \ref{eq:Fcapbdy} 
are approximated by the gradient of the color function $\nabla \alpha$, providing
\begin{equation}
    \label{eq:Fcapsurf}
        \textbf{F}_{cap}^{CSF} = \sigma k \nabla \alpha.
\end{equation}
The interface curvature $k$ is computed as,
\begin{equation}
    \label{eq:interfaceCurvature}
    k = - \nabla \cdot \textbf{n}_{I} = - \nabla \cdot \frac{\nabla \alpha}{|\nabla \alpha|}.
\end{equation}
Closure to the system of equations is provided by the following advection equation for the color function:
\begin{equation}
\label{eq:AdvEqAlpha}
    \frac{\partial \alpha}{\partial t} + \nabla \cdot (\textbf{U} \alpha) = 0.
\end{equation}

\subsection{Numerical Implementation}
%{Schemes}Discritization and Numerical Schemes}
We use the VOF method implemented in OpenFOAM\textsuperscript{\tiny\textregistered} (\url{https://openfoam.org}) with the interFoam solver, where the domain is discritized in space using the finite volume method, using an Eulerian mesh. The field variables such as velocity $\textbf{U}$, pressure $p$ and color function $\alpha$ are stored at the cell centers. Time is discritized by first order {Euler} scheme. The gradient terms are discritized using {Gauss linear} scheme, of second order accuracy. The advection term in the Navier-Stokes equation (second term in the left hand side of Eq. \eqref{eq:momBal}) is solved using the {limited linear difference} scheme. As the color function is required to be bounded, {vanLeer} scheme \citep{van1979towards} is used for the advection of the color function in Eq. \eqref{eq:AdvEqAlpha}. As many other numerical schemes, VOF suffers from numerical diffusion and smearing of interfaces, that arises from solving the discretized advection equation of the color function. To reduce this numerical artefact, an additional term $\alpha (1-\alpha) \textbf{U}_{r}$ is added to Eq. \eqref{eq:AdvEqAlpha}, where the so-called compression velocity is approximated as $\textbf{U}_{r} = \min[c_{\alpha}\textbf{U}_{cv}, \max(\textbf{U}_{d})]$ \citep{rusche2003computational}. $c_{\alpha}$ is the compression coefficient set to one based on the studies of \citet{deshpande2012evaluating,hoang2013benchmark,ferrari2013direct}, $\textbf{U}_{cv}$ refers to the velocity in a specific control volume and max($\textbf{U}_{d}$) refers to the maximum velocity in the entire porous medium. The term $\alpha (1-\alpha)$ restricts the compression velocity to act only at the interfaces.

%\footnote{define any parameter you present; here, $\textbf{U}_{cell}, \textbf{U}_{domain}$; also, use the right type--bold only for vector (eg I changed to scalar "${U} = 70 ~\textrm{mm/s}$". DONE [SP]}

The pressure-velocity coupling in the Navier-Stokes equations are solved using the Pressure Implicit with Splitting of Operators (PISO) algorithm \citep{issa1986solution}. 
%\sout{The cell size used in our simulations is $\Delta x = 13.5~\mu\textrm{m}$.}
To generate the discretized pore space for the simulations, we first use a rectangular mesh with cell size of $\Delta x = \Delta y = 13.5~\mu\textrm{m}$, and then re-meshed using the snappyHexMesh library of OpenFOAM\textsuperscript{\tiny\textregistered}. 
%\footnote{[RH: cell size = ?? mesh in b does not look rectangular!] \textcolor{brown}{[SP:] addressed now in text. We initially have Cartesian mesh. The cells are later 'snapped' (remeshed) using the 'snappyHexMesh' library of OF.}}
The time step size is chosen based on the Brackbill number $t_{Bk} = \sqrt{{\rho_{avg}{\Delta x}^{3}}/{(\pi \sigma})}$ where $\rho_{avg}$ is the average density of fluids in the domain and $\Delta x = 13.5~\mu\textrm{m}$ is the cell size \citep{brackbill1992continuum}. 
To reduce computational runtime, we set the time step size $\Delta t \approx 3 t_{Bk}$, which for our settings provides $\Delta t = 10~\mu\textrm{s}$. 
The parameters describing the physical properties are provided in Section \ref{paramSpace}.

\subsection{Settings and parameter Values}
\label{paramSpace}
% \subsection{Porous Medium}
% \label{superPM}

We simulate displacement in a geologically-realistic medium, obtained from a 2-D cross section of micro-CT image of a sand pack \citep{pak2023effects} {Fig. \ref{fig:poreMesh}. 
While our 2-D model geometry can represent sub-pore pore filling mechanisms in the $X-Y$ plane, such as film flow along solid surfaces of the particles (filling the entire gap thickness in the $Z$ direction), it cannot represent flows in the out-of-plane ($Z$) direction, where only part of the gap is filled. Thus, our model excludes corner flows \citep{blunt2001flow,mazouchi2001thermocapillary} and wetting layer formation along the top and bottom plates. 
These mechanisms become important for the invasion of strongly wetting fluid (``strong imbibition'') and strongly non-wetting fluids (``strong drainage'', at high $Ca$) at $M<<1$
%, which has been shown to be important during strong imbibition scenarios 
\citep{zhao2016wettability, primkulov2021wettability}. 
%%% We note that by considering a 2-D domain, our simulations cannot capture 3-D invasion mechanisms where the invading fluid does not fill the entire third dimension \sout{such as} \textcolor{brown}{due to} corner flows and film flows along the top and bottom boundaries. Such mechanisms become important for the invasion of strongly wetting fluid (``strong imbibition'') and strongly non-wetting fluids (``strong drainage'', at high $Ca$) at $M<<1$. 
The simulated sample {has a porosity of $\phi = 27.8\%$ and} dimensions are 6.75 mm and 14.05 mm in the $X$ and $Y$ direction. It is discretized into 145000 cells using the snappyHexMesh library of OpenFOAM\textsuperscript{\tiny\textregistered}. 
The zoom-in shows the intricate nature of the geological media, with highly nonuniform pore bodies and throats, and dead ends (Fig. \ref{fig:poreMesh}b). 
The mesh contains both Cartesian and non-Cartesian cells; non-Cartesian cells are used close to the boundaries with the solid grains, to capture the orientation of the pore spaces.

\begin{figure}[h]
    \centering
    \includegraphics[width=.8\textwidth]{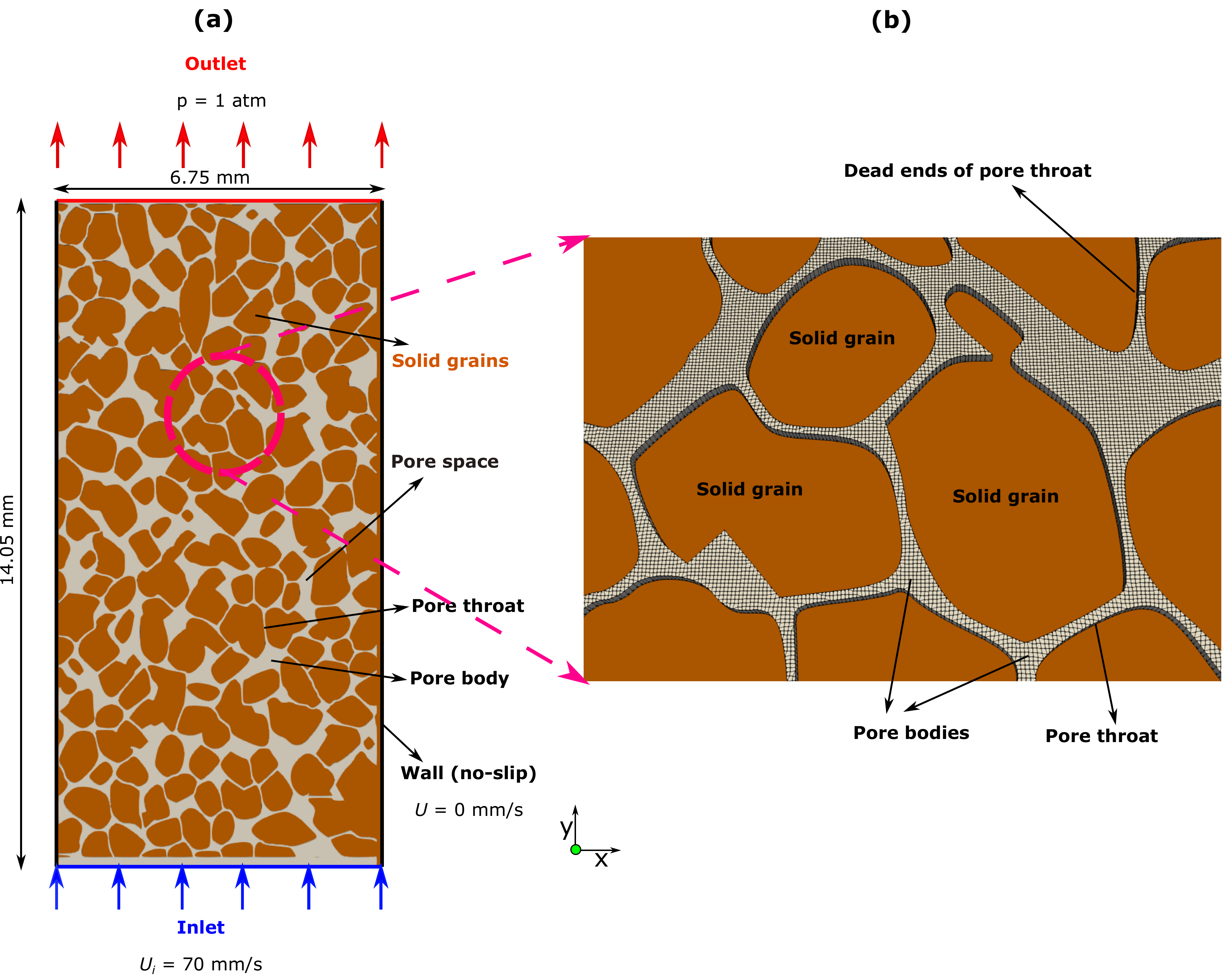}
    \caption{(a) the simulated two-dimensional porous medium. The pore space and solid grains are shown in grey and brown, respectively. 
    % pore spaces comprising of pore throats and pore bodies  
    % RH: CONFUSING! PNM has this distinnction, however DNS does not and considers all as same pore space..
    %The inlet and outlet boundary conditions used for the numerical setup are specified. 
    (b) zoom-in showing the meshed pore space.
    %  \hl{[RH: fig: what is grad(Q)? I don't think this has a physical meaning.. grad(p) does..]}
    }
    \label{fig:poreMesh}
\end{figure}

The boundary conditions are fixed injection velocity of the invading fluid at the inlet face at ${U_i} = 70 ~\textrm{mm/s}$, fixed pressure $p=1$ atm at the outlet, and no-flow at the other two (lateral) boundaries (Fig. \ref{fig:poreMesh}a). 
%The invading fluid is injected at a fixed velocity ${U} = 70 ~\textrm{mm/s}$, and the pressure at the outlet is set to $p = 1~\textrm{atm}$.
%%% \sout{Pressure boundary conditions are zero gradient at the inlet boundary and $1 ~\textrm{atm}$ at the outlet.}
%\hl{I do not understand. zero gradient at a point (or boundary) means there's no flow at that point. perhaps you are confusing zero boundary with free flow BCs? [SP:] This is how OF defines the bc. Pressure is computed at cell centres. So, we have boundary cell values for pressure. This value is interpolated to the boundary face with a zero gradient. This pressure data is then used to compute fluxes.}
%%% \footnote{RH: zero gradient at a point (or boundary) means there's no flow at that point. perhaps you are confusing zero boundary with free flow BCs? RH: still unclear. \textcolor{brown}{[SP:] to avoid confusion we now mention the injection velocity and outlet pressure.}}
%
%The other two boundaries (left and right in Fig. \ref{fig:poreMesh}a) are considered impermeable walls. 
%These walls as well as 
The surfaces of solid grains %within the medium 
are subjected to no-slip boundary conditions. The interface normal $\textbf{n}_{I}$ pointing towards the invading phase orients the interface according to 
% the following condition
% \begin{equation}
%     \label{eq:ca}
$    \textbf{n}_{I} = \textbf{n}_{w} cos(\theta) + \textbf{t}_{w} sin(\theta)$,
%\end{equation}
where $\textbf{n}_{w}$ is the normal to the solid grain surfaces, $\textbf{t}_{w}$ is the tangent to the solid grain surfaces and $\theta$ is the equilibrium contact angle that the injected phase makes with the solid surfaces \citep{brackbill1992continuum}.

We vary the contact angle in the range of $\theta~\in~[0^{\circ},180^{\circ}]$ with $\theta=20^{\circ}$ increments. Here, $\theta$ refers to the equilibrium contact angle that the injected fluid makes with the solid surfaces. 
Imbibition and drainage refer to $\theta<90^{\circ}$ and $\theta > 90^{\circ}$, respectively.
At $\theta=0^{\circ}$ (``strong imbibition''), the injected fluid perfectly wets the solid surfaces. 
At $\theta=180^{\circ}$ (``strong drainage''), the injected fluid is perfectly nonwetting, such that it repels from the solid surfaces. 
The viscosity ratio, $M = {\mu_{i}}/{\mu_{d}}$, was varied between 0.01 and 100, by setting the maximum viscosity of one fluid to $\mu_{max} = 0.1~{\textrm{kg}}/{\textrm{ms}}$ and tuning the viscosity of the other.
Here $\mu_{i}$ and $\mu_{d}$ are the viscosities of the invading and defending fluids, respectively.
%to obtain the desired $M$.
%
We set the density of both fluids to $\rho=\rho_{i}=\rho_{d}=1000~{\textrm{kg}}/{\textrm{m}^{3}}$, and the surface tension to $\sigma=0.07~{\textrm{kg}}/{\textrm{s}^{2}}$.
The capillary number, defined here as
% \begin{equation}
% \label{eq:canum}
%$    Ca = \frac{U_{i} \mu_{max}}{\sigma}$, 
$    Ca = {U_{i} \mu_{max}}/{\sigma}$
%%% \footnote{here and elsewhere, I suggest to avoid inline fractions, they come out too small (I thought I changed all of those, perhaps you added new ones); e.g. here use $    Ca = {U_{i} \mu_{max}}/{\sigma}$ \textcolor{brown}{[SP: Done]}}
% \end{equation}
was fixed at $10^{-3}$. 
%Here $U_{i}$ is the magnitude of the injection velocity. 
%\textcolor{brown}{
We chose this $Ca$ value in order to focus on viscous-dominated flow, vs. the capillary-dominated flow at $Ca \leq 1 \times 10^{-4}$ investigated elsewhere \citep{lenormand1988numerical, ferer2004crossover, an2020transition}.   
%}
%and $\mu_{max} = \max(\mu_{n}, \mu_{w})$ is the maximum viscosity of the two fluids. 
The total number of simulations were 90: 10 values of $\theta$ and 9 values of $M$.
Simulations were run using parallel computations with 16 Intel Xeon E5-2690 processors (clock speed 2.60 GHz). With that, simulating 1 physical second of flow requires runtime of $\sim$3.5 hours.

\subsection{Image processing for quantitative analysis of patterns}
\label{ImageAnalysis}

Quantitative analysis of the observed patterns at breakthrough is done using two characteristics:
(i) the displacement efficiency $D_{e}$, which is the volume of the displaced defending fluid normalized by the total pore volume; and (ii) the fractal dimensions $D_{f}$, an estimation of the roughness of the interface, which we compute using the box-counting method \citep{feder2013fractals}. The fractal dimensions in 2-D is bounded at $D_{f} \in [1 - 2]$, where $D_{f} = 1$ and $D_{f} = 2$ represent the highest possible roughness and a completely compact interfacial {morphology}, respectively.  
These computations are done on a binary format (white for the injected fluid, black for everything else). Conversion of the invasion images into binary format was done using Fiji software \citep{schindelin2012fiji}.
%
% Later, we select all the ganglions that exist in the porous medium at the time of breakthrough and overlay the internal spaces within a ganglion with white color to capture exclusively the boundaries of all the ganglions.

%-------------------%
\section{Results}
\label{Results}
% Insights from 90 Direct Numerical Simulations performed at different contact angles and fluid viscosities are summarized in this section. The investigated parameter space is discussed in Section \ref{paramSpace}. 
% Flow regimes have been classified based on visual observations and fractal dimensions. We further discuss insights from stability analysis that can assist in approximately delineating the crossover between flow regimes. 
% We finally discuss the pore filling mechanisms observed for different scenarios.

\subsection{Displacement Patterns}
%\subsection{Fluid Displacement Patterns as a Function of Wettability and Different Fluid Viscosities}
\label{FluidDisplacementPatterns}

We begin with a qualitative analysis based on the visual appearance of the patterns, followed by quantitative characterization using fractal dimension. 
Figure \ref{fig:invasionPatterns} shows the simulated patterns at breakthrough for the 90 conditions (varying independently $M$ and $\theta$) considered here. 
The displacement patterns change from viscous fingering (VF) to compact displacement (CD), with {an intermediate regime} (VF/CD) exhibiting a mix of features from both. 
%Viscous fingering is seen to occur in drainage at low $M$ and in stable displacements can be seen at large $M$ irrespective of the wettability conditions that exist.
VF, characterized by long and narrow fingers, emerges at $M \leq 0.01$ irrespective of $\theta$. 
This was also observed in \cite{primkulov2021wettability}, however at $M < 0.5$.
At $M \geq 10$, the invading fluid fills most of the pore space leading to CD, again irrespective of $\theta$.
At intermediate values of $M \approx 1$, a transition between VF and CD occurs; the $M$ value for the transition among the regimes depends on $\theta$. 
The fluid fingers in imbibition are slightly wider compared to those in drainage, also observed in \cite{mora2021influence}. This is due to the increased tendency of the invading fluid to minimize contact with the solid surfaces in drainage.
%
%Similar observations however at slightly different $M$ values were observed in \cite{primkulov2021wettability}: VF at $M < 0.5$, and 

\begin{figure}[!]
    \centering
    \includegraphics[width=.81\textwidth]{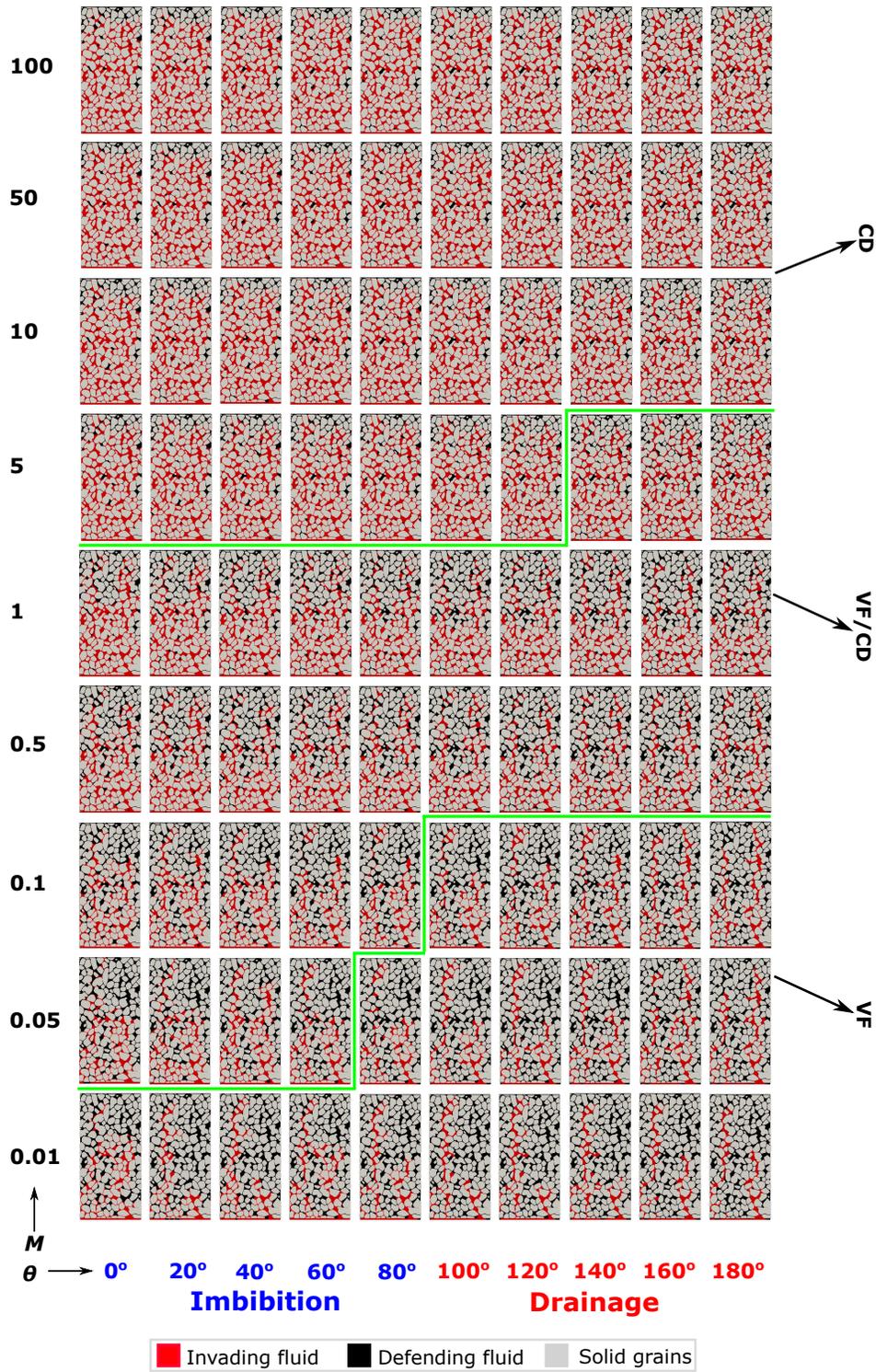}
    \caption{Fluid displacement patterns for different contact angles $\theta$ and viscosity ratios $M$. The invading fluid is shown in red, defending fluid in black and solid grains in grey. 
    % Imbibition occurs when the contact angle is below $\theta < 90^{\circ}$. 
    % Drainage occurs when the contact angle is greater than $\theta > 90^{\circ}$. 
    The continuous green lines indicate transition between flow regimes: viscous fingering (VF) to intermediate (VF/CD) and intermediate to compact (CD). 
    % \textcolor{brown}{[SP:] (A) This is quite a work to change. We are thinking retain these as they are for now and wait for the reviewers comments. (B) Done}
    % \hl{[RH: 
    % (A) %SEE COMMENT ON CLASSIFICATION, AND HOW TO PRESENT PATTERNS (IE SHOW THE 2ND FLUID IN ANOTHER COLOR?).
    % Graphics: Better, but still, the top ones don't "feel" compact to the naked eye, ie they looks very patch/spongy/CF-like (maybe because white somehow seems like void space?). If we want to try improve, one idea you can try quickly is coloring solid with a similar (e.g. slightly more pale pink) than invading fluid; this way it will give a feel of compact which covers (almost) the entire space. Technically, to avoid the work of multiple reformatting of ALL patterns, try for a single compact pattern (if you want, we can discuss back and forth few versions by email or skype), and once it works change all boxes.
    % (B) FORMATTING: Small issues: typo "Cpmpact"; also consider using the acronyms VF, CD and VF/CD instead. 
    % %I suggest to (1) remove outer "phase boundaries" (unnecessary and even wrong as VF and CD would occur at much lower and higher M, respectively); (2) instead, only show 2 boundaries inside the fig: a. between VF and transition ; b. between transition and CD. (3) Make boundaries more visible, e.g. thicker or more visible color. (4) keep the text of the 3 regimes, could use black instead of colors.
    % ]}
    }
    \label{fig:invasionPatterns}
\end{figure}

Supporting material provides three videos of the invasion processes occurring at $\theta = 60^{\circ}$ for $M = 0.01$ (VF, Video 1), 1 (VF/CD, Video 2) and 100 (CD, Video 3). 
For $M = 1, \theta = 60^{\circ}$ (Video 2), we notice the developed fingers propagating ahead of the compact displacement front. This displacement pattern eventually results in showcasing the traits of both VF (towards the outlet) and CD (towards the inlet).

Next, we analyze the patterns quantitatively, using (a) the displacement efficiency $D_{e}$; and (b) the fractal dimensions $D_{f}$.
%We use two characteristics to analyse the patterns in Fig. \ref{fig:invasionPatterns} quantitatively: 
%(i) the displacement efficiency $D_{e}$, which is the volume of the displaced defending fluid normalized by the total pore volume (Fig. \ref{fig:de_df_pd}a); and (ii) the fractal dimensions $D_{f}$, an estimation of the roughness of the interface computed from the box-counting method \citep{feder2013fractals} (Fig. \ref{fig:de_df_pd}b). Conversion of the images in Fig. \ref{fig:invasionPatterns} into a binary format (white for the injected fluid, black for everything else) was done using Fiji software \citep{schindelin2012fiji}.
%
The more compact, less preferential invasion is characterized by larger $D_e$ and $D_f$  (Fig.~\ref{fig:de_df_pd}). Consequently, as the invading fluid becomes more wetting i.e. as $\theta$ decreases, both $D_e$ and $D_f$ increase, in most cases regardless of $M$. Similarly, for a given $\theta$, increasing $M$ stabilizes the displacement thus increasing $D_e$ and $D_f$. The efficiency increases from $D_e < 30$\% at $M=0.01$ and $\theta = 180^\circ$ (drainage at non-favorable viscosity ratio, VF) to $D_e > 80$\% at $M=100$ and $\theta = 0^\circ$ (imbibition at favorable viscosities, CD). 
%
%\cite{lovoll2004growth,toussaint2005influence,holtzman2010crossover,islam2014characterization,zhao2016wettability,chen2017visualizing} NO NEED FOR ALL THESE REFS!!

%
% $D_{e}$ is simply the volume of the displaced defending fluid normalized by the total pore volume. 
% $D_{f}$ is computed from the box-counting method \citep{feder2013fractals}.

% Based on the data related to the invasion patterns presented in Fig. \ref{fig:invasionPatterns}, we discuss the trends noticed for two metrics namely, 1. displacement efficiency $D_{e}$, and, 2. fractal dimensions $D_{f}$.  
% The displacement efficiency $D_{e}$ indicates the ratio of the cumulative volume of the defending fluid that has been displaced out of the porous medium until breakthrough to the cumulative pore volume of the the porous medium. We commence our simulations with the defending fluid occupying the entire pore spaces. Therefore, we compute the cumulative volume of the injected phase in the porous medium until breakthrough to determine the volume of defending fluid displaced. The displacement efficiencies for all the investigated cases are shown in Fig. \ref{fig:de_df_pd}a.

\begin{figure}
    \centering
    \includegraphics[scale=0.5]{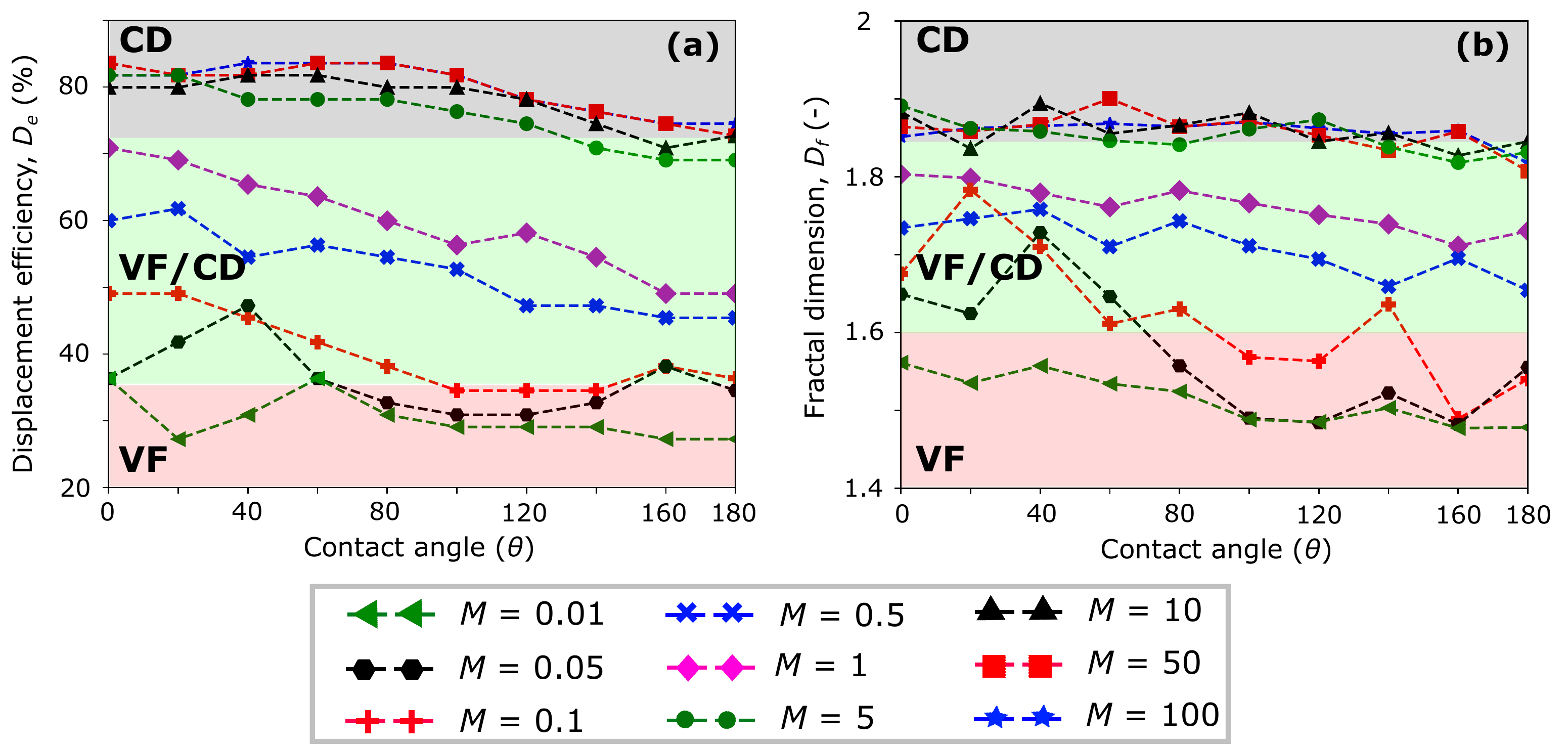}
    \caption{Quantitative analysis of the displacement patterns in Fig. \ref{fig:invasionPatterns} using (a) displacement efficiency $D_{e}$; and (b) fractal dimension $D_{f}$. In panels, the background colors of red, green and grey refer to the range of $D_{e}$ and $D_{f}$ observed for the different regimes based on a combination of visual appearance and stability analysis. %\textcolor{brown}{[SP: Done]}
    %: 
    %viscous fingering ($D_{f} \leq 1.6$), transition flow regime ($1.6 < D_{f} \leq 1.83$) and compact displacements ($D_{f} \geq 1.83$). 
%    \hl{[RH: Formatting: (1) avoid "double legend", remove the lower onem replace with text labels "CF", "VF" etc on shaded regions. (2) RHS panel: letter (b) missing under shading; (c) on both use smaller fonts for a, b.
    %did you try to put the shading (for various regimes) on panel a? it is a bit weird we only use Fd for that. also, format comment: make the shading colors in b weaker, they will be still be easily visible and the points will be much more visible. [SP:] (Done!!)
%    ]}
    %The ranges for the fractal dimension values indicating the crossover between different flow regimes is based on the phase diagram and stability analysis presented in Fig. \ref{fig:phaseDia}.  
    }
    \label{fig:de_df_pd}
\end{figure}

Combining the qualitative classification of the 90 simulated patterns (based on visual appearance, cf. Fig.~\ref{fig:invasionPatterns}) with their $D_{f}$ values, allows us to establish the corresponding range of $D_f$ values for each regime: $D_{f} <1.6$ for VF, $D_{f}=1.6-1.83$ for intermediate (VF/CD) regime, and $D_{f}>1.83$ for CD; our $D_f$ values agree well with published values for these regimes \citep{lovoll2004growth}. 
%While the evaluation of $D_f$ has appreciable uncertainty, the values of $D_f$ obtained here are in overall agreement with published values for VF, $D_{f} < 1.6$ and CD, $D_{f} > 1.83$ \citep{lovoll2004growth}.
Plotting these in the form of a phase diagram, provides a quantitative estimation of the phase boundaries between regimes (dashed green line in Fig.~\ref{fig:phaseDia}). 
These phase boundaries (dashed green line) agree well with 
%Next, we compare these phase boundaries (dashed green line) with 
%This provides a threshold $M$ values corresponding to the various regimes (greed dotted lines). 
%
%These boundaries, namely the threshold $M$ values, agree well with
%\footnote{\hl{[RH: I would avoid (here and elsewhere) the term "threshold $M$", because it imply a sharp transition or emergence of different physics where there is no critical transition, it is rather smooth with wide transition regime.] [SP: Agree. We will anyway discuss this topic later in the text.]}}
theoretical values obtained using the classical stability analysis by \citet{saffman1958penetration} (plotted as green dots in Fig.~\ref{fig:phaseDia}); for derivation details see Appendix \ref{appendix_A}.
% RH: TOO MUCH DETAILS WHICH ARE DISTRACTING FROM OUR STORY; THEREFORE I PUT IT AS APPENDIX
%
The value of $M$ at the boundary between regimes increases with $\theta$, in particular for the crossover between VF and VF/CD (Fig. \ref{fig:phaseDia}). 

\begin{figure}[!]
    \centering
    \includegraphics[width=.6\textwidth]{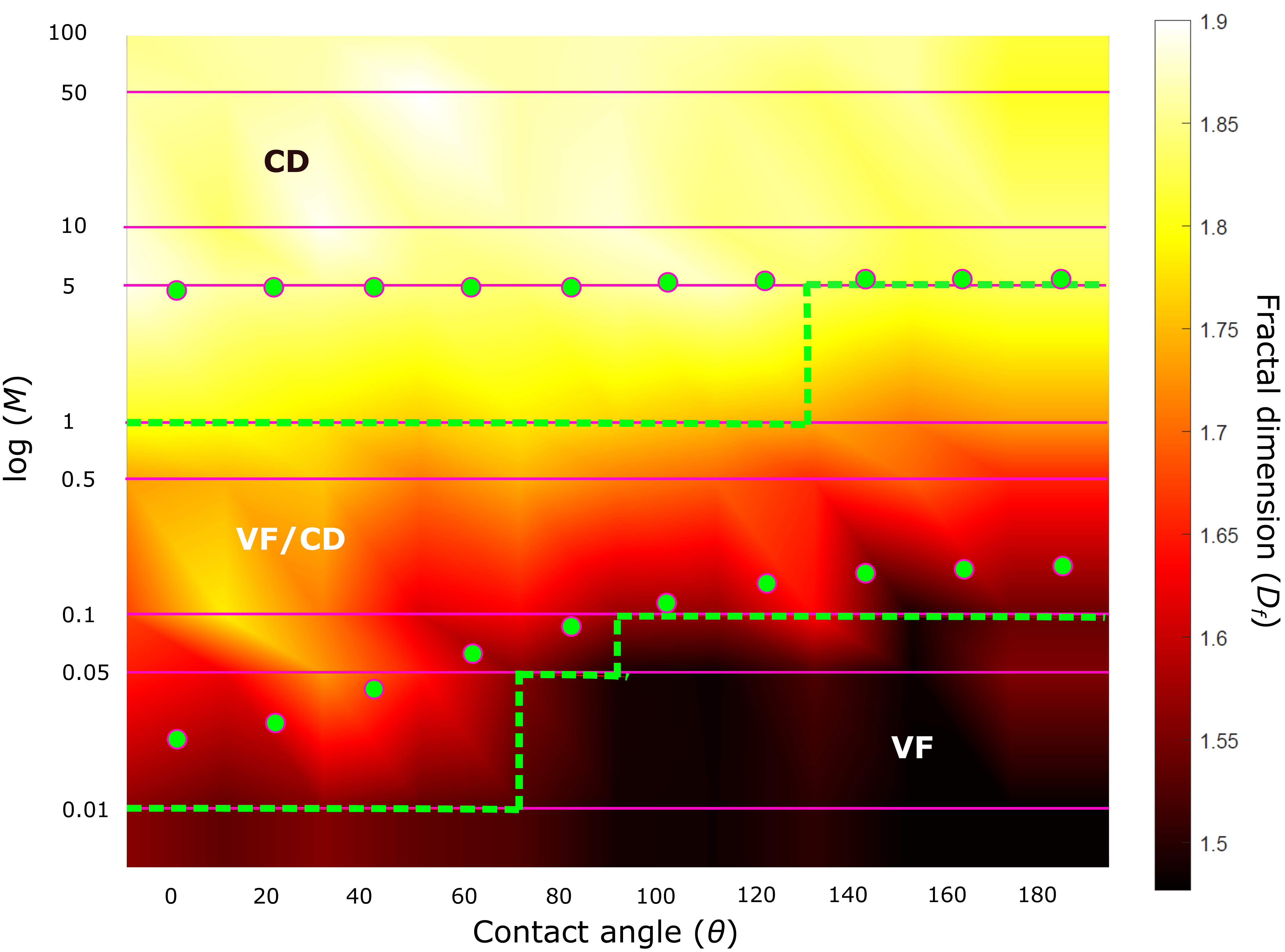}
    \caption{
    Phase diagram of displacement patterns classified based on $D_{f}$, as a function of viscosity ratio $M$ and wettability $\theta$. 
    Results of $D_{f}$ from simulations (90 combinations of $M$ and $\theta$) are linearly interpolated. 
    Dashed green lines indicate boundaries between regimes 
    %(viscous fingering to intermediate regime and intermediate regime to compact displacement) 
    based on visual observations and $D_{f}$.
    The green dots indicate theoretical phase boundaries estimated through stability analysis.
    %
    %\sout{Fractal dimension $D_{f}$, shown in colorbar, is in good agreement with the phase boundaries between the 3 regimes (viscous fingering, transition, and compact displacement) obtained by two different means: (i) color gradients {[RH=??? not clear; see comments in text]} and visual inspection of patterns (dashed lines); and (ii) stability analysis (dotted lines).}
    %The green dotted lines indicate the crossover between different flow regimes (viscous fingering to transition regime and transition regime to stable displacement) classified based on color gradients and visual observations. The approximate range of $D_{f}$ for various flow regimes is indicated next to the color bar. The green dots indicate the crossover between different flow regimes determined based on stability analysis.}
    %\hl{[RH: formatting: avoid pink font for regime names. use white font for VF and Intermediate regime, black for Compact.] [SP: Done.]}
    }
    \label{fig:phaseDia}
\end{figure}

\subsection{Pore Filling Mechanisms}
\label{poreFilling}

%%% \textbf{[RH: I found the new explanation unclear, hence revised thoroughly below what I could, and commented where I could not.]\\}
The pore-scale mechanisms, controlling the manner by which pores are filled, eventually dictate the larger, sample (macroscopic) scale patterns. 
Valuable information about these mechanisms is obtained here by analysing the pore size distribution (PSD) of the invaded pores. 
The pore sizes are determined using the distance transform watershed method, according to the maximum diameter of an inscribed circle $p_{s}$ which fits in them \citep{legland2016morpholibj}. We classify pores into 3 size groups: small for $p_{s} <= 0.1~\textrm{mm}$, medium for $0.1~\textrm{mm} > p_{s} \leq 0.2~\textrm{mm}$ and large for $0.2~\textrm{mm} > p_{s} \leq 0.3~\textrm{mm}$.
The number of invaded pores (normalized by the total number of invaded pores) $N_{i}^{*}$ for different pore size distributions and for nine $M, \theta$ combinations is shown in Fig. \ref{fig:poreNumber}. 
%We divide the number of invaded pores of different sizes with the total number of invaded pores  to compute $N_{i}^{*}$.
%\sout{, normalized by ... [see comment on analysis, either change it or explain better what you did].}
%{Here, $N_{i}^{*} = {N_{i}}/{\max(N_{i} (p_{s}))}$ where $\max(N_{i} (p_{s}))$ refers to the maximum number of pores occupied by the invading fluid for a that specific $M, \theta$ combination, for the 3 pore sizes categories $p_{s}$. In Fig. \ref{fig:poreNumber} $\max(N_{i} (p_{s}))$ are highlighted by (*). [??? unclear]}
%\footnote{[RH: Analysis: what are you normalising by (what is the meaning of this maximum you use now $\max(N_{i} (p_{s}))$)? why not simply the total number of invaded pores, so that the sum of the Ni* of 3 groups is 1?]\textcolor{brown}{[SP: To avoid this confusion, we no longer normalize. We use the original data as it is.]}}
%
This demonstrates the strong effect of wettability at small (unstable) viscosity ratio, $M = 0.01$ (VF; Fig. \ref{fig:poreNumber}a). 
In contrast, it shows the small effect of $\theta$ in stable, compact displacement, with $M = 100$ (CD; Fig. \ref{fig:poreNumber}c). 
It is also interesting to examine the PSD symmetry: the invaded PSD is relatively symmetric (and uniform) for $M = 100$, {slightly} skewed for $M = 1$, and strongly skewed for $M = 0.01$ at strong imbibition ($\theta = 0^\circ$) (Fig. \ref{fig:poreNumber}). 
%
%For unstable displacement $M = 0.01$, there is strong skewness for strong imbibition ($\theta = 0^\circ$).
%
To explain this, we further analyse the pore filling mechanisms during individual invasion events.

\begin{figure}
    \centering
    \includegraphics[scale=0.35]{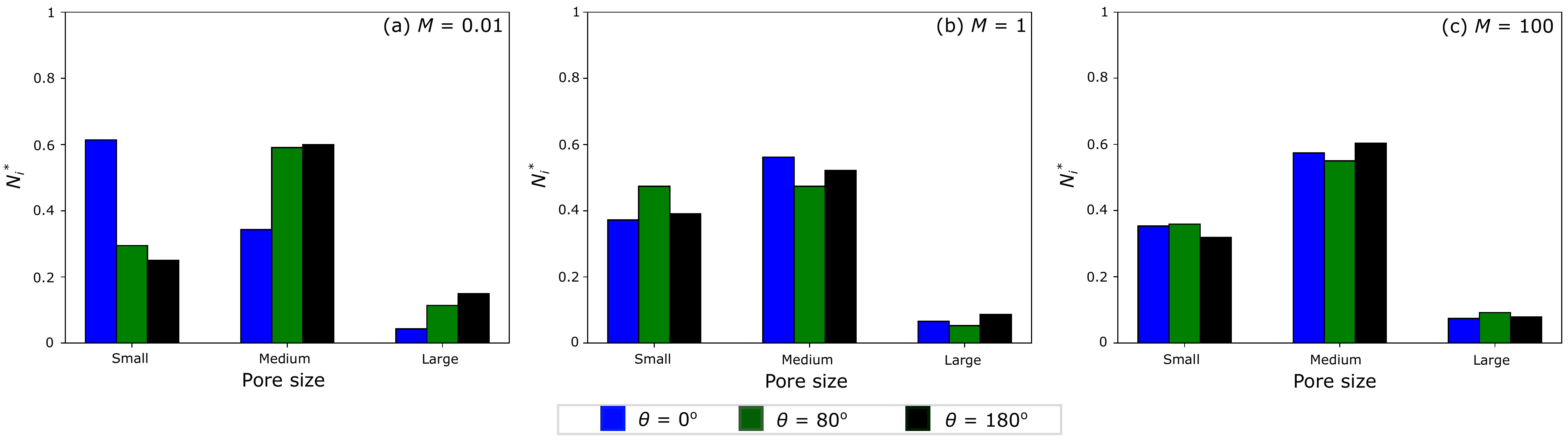}
    \caption{
    % ORIGINAL TEXT
    %Normalised number of pore spaces of different sizes occupied by the injected fluid at contact angles $\theta = 0^{\circ}$ (strong imbibition), $\theta = 80^{\circ}$ (weak imbibition) and $\theta = 180^{\circ}$ (strong drainage) at viscosity ratios (a) $M = 100$ (flow regime: stable displacement), (b) $M = 1$ (flow regime: transition flow regime) and (c) $M = 0.01$ (flow regime: viscous fingering). We take the ratio of the number of pore bodies occupied by the invading fluid to the maximum number of pore spaces occupied by the invading fluid for each case respectively to obtain the normalised number of pore spaces occupied by the invading fluid.
    Normalized invaded pore size distribution for (a) $M$ = 0.01 (VF), (b) $M$ = 1 (VF/CD) and (c) $M$ = 100 (CD) at $\theta = 0^{\circ}, 80^{\circ}, 180^{\circ}$. Small pore size refer to $p_{s} <= 0.1~\textrm{mm}$, medium for $0.1~\textrm{mm} > p_{s} \leq 0.2~\textrm{mm}$ and large for $0.2~\textrm{mm} > p_{s} \leq 0.3~\textrm{mm}$. 
    }
    \label{fig:poreNumber}
\end{figure}
{
We investigate the pore filling mechanisms in a small region composed of a few pores both qualitatively (visually) and quantitatively through the evolution of the local capillary pressure $p_{c}$.  
The local capillary pressure, $p_{c} = p_{n} - p_{w}$ where $p_{n}$ and $p_{w}$ are the volume-average non-wetting and wetting phase pressure, respectively, is computed as $p_{i} = \frac{\sum p_{i} V_{c}}{\sum V_{c}}$ where $i \in (w,n)$ 
%subscript $i$ refers to wetting ($w$)/ non-wetting ($n$) phase 
and $V_{c}$ is the volumes of the computational cells within the analyzed region. 
%exclusively in the region where we perform qualitatively analysis. 
We track the evolution of the local pressure $p_{c}$ vs. the local wetting phase saturation $S_{w}^{*}$. The latter is normalized by the maximum $S_{w}$ attained as the invading fluid reaches breakthrough on all four boundaries of the window of observation (red rectangle in Fig.~\ref{fig:imbibition}).
%) domain of inspection used for qualitative analysis.
%We plot $p_{c}$ against $S_{w}^{*}$. $S_{w}^{*}$ refers to the normalized wetting phase saturation computed by dividing $S_{w}$ during various stages of the pore filling event to the maximum $S_{w}$ attained as the invading fluid reaches breakthrough on all four boundaries of the rectangular domain of inspection used for qualitative analysis.
}
In imbibition, we find that pores are filled primarily by two mechanisms: (i) film flow and (ii) cooperative pore filling, depending on the viscosities (Fig.~\ref{fig:imbibition}).
{At $M = 0.01$ (VF), the wetting phase advances as thin films coating the solid surfaces \citep{zhao2016wettability} (Fig. \ref{fig:imbibition}b). With this mechanism, the invasion progresses predominantly through smaller pores, in accordance with the PSD analysis (Fig. \ref{fig:poreNumber}a.)} 
The evolution of $p_{c}$ shows a decreases in $p_{c}$ magnitude upon imbibition (i.e. increasing $S_{w}^{*}$), see Fig. \ref{fig:imbibition}d, 
which can be due to the formation of wetting layers that eventually results in the formation of smaller interfacial curvatures $p_{c} \propto k$. 
%is shown in Fig. \ref{fig:imbibition}d. We notice that the $p_{c}$ decreases in magnitude upon imbibition (increase in $S_{w}^{*}$). This observation can be explained by the formation of a number of wetting layers that eventually results in the formation of smaller interfacial curvatures $p_{c} \propto k$. 
At moderately-high $M$ of $1 - 100$, cooperative filling becomes the dominant mechanism (Fig. \ref{fig:imbibition}c), resulting in a more stable, compact front. This mechanism fills the various pore sizes more uniformly than at $M = 0.01$, as can be seen by comparing the PSD of filled pores in Fig. \ref{fig:poreNumber}b--c (moderately-high $M$) vs Fig. \ref{fig:imbibition}a ($M = 0.01$).
The local pressure $p_{c}$ drops (by $\sim$ 500 {Pa}) 
%to $p_{c} = 0 ~\textrm{Pa}$ 
until $S_{w}^{*} \approx 0.3$ as the invading fluid reaches the entrances of the pore body. As the invasion continues, 
%At $S_{w}^{*} > 0.3$, 
several interfaces merge which increases $p_{c}$ at $S_{w}^{*} = 0.4$. Following that, cooperative pore filling continues and 
%As imbibition further progresses with cooperative pore filling, 
$p_{c}$ reaches a steady value 
%equilibrium value 
as the average menisci curvature during these events does not change significantly.    

\begin{figure}
    \centering
    \includegraphics[scale=0.32]{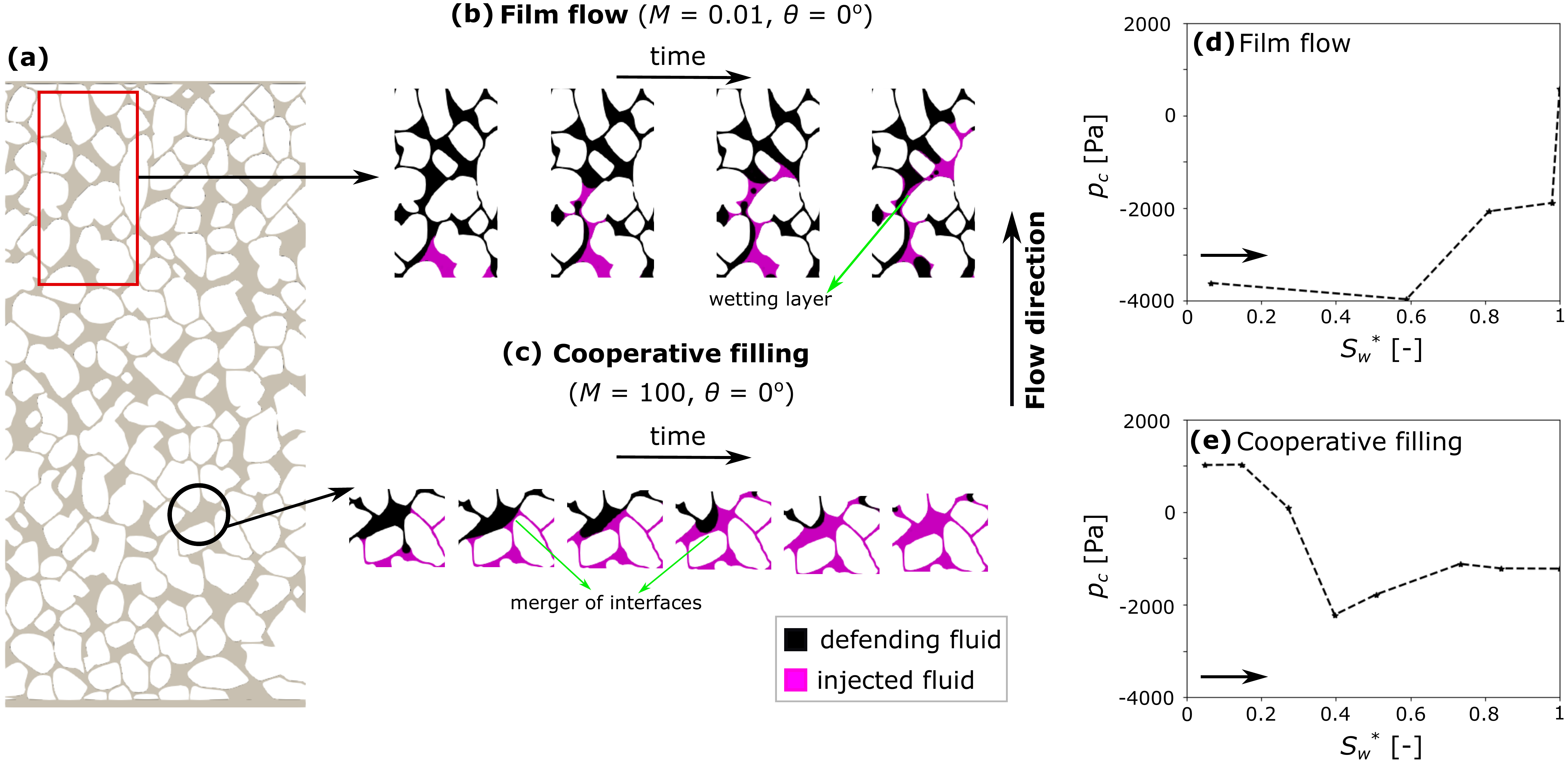}
    \caption{ 
    % ORIGINAL TEXT: Pore filling mechanisms observed during imbibition. Successive snapshots of preferential imbibition observed at $M = 0.01, \theta = 0^{\circ}$ where the injected fluid (in pink) invades smaller pore spaces is shown at the top. A signature of this invasion mechanism is the propagation of the wetting fluid in the form of wetting layers that coat the surface of the solid grains is highlighted. Successive snapshots of a pore body connected to seven pore throats subjected to co-operative pore body filling observed at $M = 100, \theta = 0^{\circ}$ is shown at the bottom. A signature of this pore filling mechanism is the gradual movement of the interfaces upon successful mergers as highlighted. The evolution of the dynamic capillary pressure against normalised water saturation ($S_{w}^{*}$) is shown for the two pore filling mechanisms. The water saturation is normalised by dividing the water saturation with the maximum water saturation attained at the end of the event. The arrows shown in the capillary pressure figure indicate the direction in which the water saturation moves.
    %
    Pore filling mechanisms in strong imbibition ($\theta=0^\circ$). Focusing on small region containing several pores (red rectangle in panel a), progression of invasion is shown as successive snapshots (injected fluid in pink, defending fluids in black and solid in white), for unfavorable and favorable $M$, 0.01 and 100, respectively (panels b, c). 
    At $M = 0.01$, the injected fluid propagates in the form of wetting layers that coat the solid surfaces (green arrow), invades smaller pores (b).
    At $M = 100$, cooperative pore filling (green arrows) is the dominant mechanism, leading to more uniform invasion of both small and large pore (c). 
    Panels d--e show the evolution of the local capillary pressure with saturation of wetting fluid ($S_{w}^{*}$) for the two mechanisms (arrows indicate direction of change in $S_{w}^{*}$ with time).
}
    \label{fig:imbibition}
\end{figure}
%\hl{[RH: Better now. As said above, unclear if Pc analysis helps. Formatting: numbering of the panels is awkward: use b , c for the two mechanisms, and if we keep pc use d, e for these two.]}
%FORMATTING: (1) label panels with a,b,c?; (2) what is "preferential flows / imbibition"? this is not a conventional nomenclature. Did you mean "film flow"? please correct everywhere incl figs; (3) replace "capillary pressure" in axis with $p_c$.] [SP:] Done.

For strong drainage ($\theta = 180^{\circ}$), the two dominant pore filling mechanisms are (i) {intermittent local jumps (``bursts'')} and (ii) cooperative filling (Fig. \ref{fig:drainage}).
At $M = 0.01$, the pores are filled by a sequence of localized bursts, leading to VF (Fig. \ref{fig:drainage}b). As in this regime viscous forces dominate over capillary forces, the location of invasion depends more on the global pressure gradients controlled by pore connectivity and less on the local pore sizes.
%\textcolor{brown}
{
This is why the PSD in this case is relatively uniform (Fig. \ref{fig:poreNumber}a).
%Hence in Fig. \ref{fig:poreNumber}a, we notice a combination of small--medium--large pores being invaded. 
%Note at $M= 0.01$ for $\theta = 0^{\circ}, 80^{\circ}, 180^{\circ}$, equal numbers of medium sized pores are invaded. However, there exists a substantial differences in the number of the small sized pores occupied by the invading phase. 
The local pressure $p_{c}$ decreases as drainage progresses (decrease in $S_{w}^{*}$; Fig. \ref{fig:drainage}d). This can be explained due to the formation of menisci that remain stagnant % from the time they form, 
due to interfacial readjustments 
%few interfaces that reconfigure (suck back) and due to 
and the evolution of the curvatures of the invading fluid front. 
%\footnote{RH: unclear. In general I am still not convinced that adding Pc analysis here adds insight, or diffuses the message and confuses..}  
}
As $M$ increases, cooperative filling becomes dominant, making the displacement pattern more compact (Fig. \ref{fig:drainage}d), and increasing the uniformity of invaded pore sizes (\ref{fig:poreNumber}b--c). 
While the mechanism seems similar to that in imbibition at high $M$, in drainage as the invading fluid is less wetting it does not completely displace the defending (wetting) fluid which remains trapped in small pockets (Fig. \ref{fig:drainage}c).
Unlike cooperative filling in imbibition where $p_{c} \approx 0$ as the interfaces reach entrance to the pore body, in drainage the tendency of the invading fluid to repel from the solid surfaces and invade mainly the larger pores results in maintaining a finite interfacial curvature at all times (vs. $p_{c}  \approx  0$ in imibibition), cf. Fig. \ref{fig:drainage}e. Here, $p_{c}$ decreases as the invading fluid fills the pore bodies ($S_{w}^{*} = 0.8 - 0.5$), with a very moderate rise after interfaces merge.
%together, we only notice a marginal rise in the overall average meniscus curvature.   

\begin{figure}[!]
    \centering
    \includegraphics[scale=0.32]{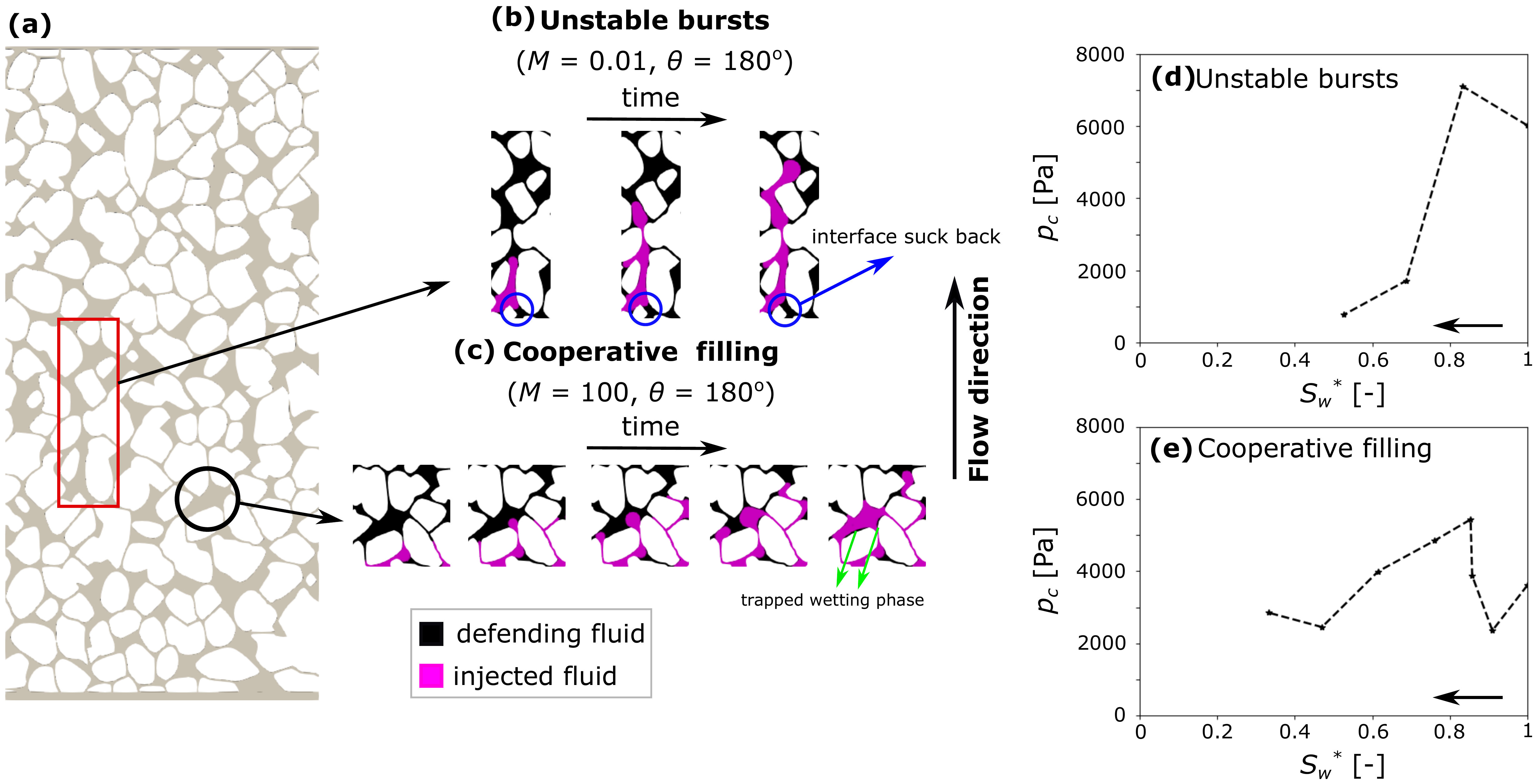}
    \caption{
    % ORIGINAL TEXT: Pore filling mechanisms observed during drainage. Successive snapshots of flow subjected to viscous instability observed at $M = 0.01, \theta = 180^{\circ}$ where the injected fluid (in pink) invades pore throats and pore bodies in secession is shown at the top. Successive snapshots of a pore body connected to seven pore throats subjected to co-operative pore body filling observed at $M = 100, \theta = 180^{\circ}$ is shown at the bottom. A signature of this pore filling mechanism is the propagation of the interfaces upon successful mergers of the nearby interfaces. The event of co-operative pore body filling during imbibition can be seen in Fig. \ref{fig:imbibition} for comparison. The evolution of the dynamic capillary pressure against normalised water saturation ($S_{w}^{*}$) is shown for the two pore filling mechanisms. The water saturation is normalised by dividing the water saturation with the maximum water saturation attained at the end of the event. The arrows shown in the capillary pressure figure indicate the direction in which the water saturation moves.
     Pore filling mechanisms in strong drainage ($\theta = 180^{\circ}$). 
     Focusing on small region containing several pores (panel a), progression of invasion is shown as successive snapshots (injected fluid in pink, defending fluids in black and solid in white), for unfavorable and favorable $M$, 0.01 and 100, respectively (panels b, c). 
    At $M = 0.01$, the injected fluid propagates by a sequence of unstable bursts, which also involve interfacial readjustments (blue arrow, panel b).
    At $M = 100$, cooperative pore filling (green arrows) is the dominant mechanism; however, unlike in imbibition, as the invading fluid is nonwetting here it leaves small pockets of trapped defencing fluid (green arrows; panel c).
    Both mechanisms 
    %\textcolor{brown}{observed during drainage} RH: NO NEED, THIS FIG IS ONLY ON DRAINAGE!
    lead to relatively uniform invasion in terms of pore sizes.
    %of \sout{both} small\textcolor{brown}{--medium--} \sout{and} large pores. \sout{{[RH: is that true? if not correct me!}]}
    Panels d--e show the evolution of the local capillary pressure with saturation of wetting fluid ($S_{w}^{*}$) for the two mechanism (arrows indicate direction of change in $S_{w}^{*}$ with time). 
    % \hl{[RH: FORMATTING: label panels with a,b,c?; again, I suggest to replace "viscous instability" with conventional nomenclature, e.g. unstable jumps (``bursts'') ? also, remove hypehn from co(-)operative. please correct everywhere incl figs]}
    }
    \label{fig:drainage}
\end{figure}
% {[RH: Better now. As said above, unclear if Pc analysis helps. Formatting: numbering of the panels is awkward: use b , c for the two mechanisms, and if we keep pc use d, e for these two.]}

%\hl{RH: I tried to revise the par. below; see if I did it right. And, I would be careful trying to say we see a different physics ("Contrarily, ..") using CFD than highly cited simulations and experiments.. [SP:] Done.}
%
The presented analysis of pore filling mechanisms provides an interesting link between the effective wettability conditions and the viscosities, $M$. 
Published simulations and experiments observed cooperative filling at intermediate wet conditions \citep{cieplak1988dynamical, cieplak1990influence, zhao2016wettability}.
Our simulations suggest that $M$ changes the effective wettability, and thus the dominant mechanism: from cooperative filling at moderate and high $M$ to film flow (during imbibition) and bursts (during drainage) at low $M$.

\section{Summary and Conclusions}
\label{conclusions}

%\textcolor{brown}
{
We leverage high-resolution Direct Numerical Simulation (DNS) to uncover the synergistic impact of wettability and viscous forces in viscous-dominated multiphase flow through a heterogeneous geologically-realistic porous media.
We present a phase diagram classifying invasion patterns into viscous fingering (VF), compact displacement (CD), and intermediate regime (CD/VF), with a transition from compact upstream to VF downstream. 
%
%\st{During VF, viscous forces of the defending fluid govern the invasion patterns, During CD, viscous forces of invading fluid dominate. During the intermediate regime, viscous forces of both fluids compete against each other.} 
%\hl{I suggest to remove the above, it is not new nor I think it helps explaining the below sentence..}
%\st{As a result, CD/VF is characterized by compact displacement near the inlet, which eventually transforms into VF as the interface traverses through the porous media. }
%
At the macroscopic (sample) scale, the wettability of porous media plays a pivotal role in controlling the crossover between regimes. Our simulations indicate an increase in threshold $M$ (at which the crossover between regimes occurs) as wetting properties vary from imbibition to drainage. 
%
%\st{As the scale of investigation reduces (flow analysis within finite number of pores), significant variations in the invasion protocol occur with the wetting conditions of porous media.} 
%
Wettability was also found to affect the pore filling mechanisms.
For 
%For example, at an extremely 
low $M$ (VF), film flow dominated during imbibition and bursts during drainage. 
%The change in invasion protocol from viscous burst to film flow signifies the minimization of viscous dissipation. 
%% RH: unclear sentence 
This strong effect of wettability
%% However, the wettability control 
over the pore-filling behavior diminishes as $M$ increases: 
%; thus, the 
cooperative filling was found to be dominating in CD 
%the compact displacement regime,
irrespective of wettability. 
This change in mechanisms indicates a change in effective wettability conditions from neutral-wet to strong-wet, respectively. 
These intriguing effects of viscosity on effective wettability should be considered in modeling multiphase fluids of similar viscosities, which is of interest to applications such as non-aqueous phase liquid (NAPLs) contamination and enhanced  hydrocarbon recovery. 

\appendix
\section{Derivations of phase boundary from linear stability analysis}
\label{appendix_A}

Here we describe the evaluation of the theoretical phase boundaries (green dots in Fig.~\ref{fig:phaseDia}) using the linear stability analysis by \citet{saffman1958penetration}. 
Considering the fluid flow potential $\phi = \nabla \textbf{U}$, the continuity equation Eq. \ref{eq:massBal} for each individual phase $i$ can be written as
\begin{equation}
    \nabla \cdot \nabla \phi_{i} = \nabla^{2} \phi_{i} = 0.
\end{equation}
With this, force balance becomes \citep{saffman1958penetration} 
\begin{equation}
    \bigg[ \textbf{U} \epsilon - \frac{\alpha \epsilon}{\gamma} \bigg] \frac{\mu_{def}}{k_{def}} - 
    \bigg[ \textbf{U} \epsilon + \frac{\alpha \epsilon}{\gamma} \bigg] \frac{\mu_{inj}}{k_{inj}} = 
    \frac{2 \sigma cos (\theta)}{r}
    \label{eq:stabilityAna}
\end{equation}
where $\epsilon$ 
%\textcolor{brown}{[SP:] (actually these are 2 different variables as per the ref. I merge them to be $\epsilon$)}
%\footnote{[RH: should be $h_\epsilon$, ie $\epsilon $ is a subscript? as it was written, it looks like a product of two variables. \textcolor{brown}{SP: in lit. these are 2 vars. In our analysis, anyways eventually they cancel out. So, I now merged these 2 vars as one $\epsilon$ in text and eqs.}.}
represents the {location of the perturbed displacement (finger) front relative to the base state---the interface morphology before fingers develop. Here, $\alpha$ is the growth rate of the perturbations and $\gamma$ is the wavenumber indicating the number of periodic disturbances in the developed finger (see further details in \citet{saffman1958penetration,rabbani2018suppressing}).
{
The effective permeability of the injected fluid can be approximated as $k_{i} = {2 \phi \lambda r \epsilon}/{n}$ where $r$ is the pore radius, $n$ are the total number of pores in the considered porous medium \citep{rabbani2018suppressing}. 
$n$ is determined from the number of pores occupied by the invading fluid $N_{i}$ in Fig. \ref{fig:poreNumber}c at $M = 100$ (CD), using the distance transform watershed method in Fiji software \citep{legland2016morpholibj}. At $\theta = 0^{\circ}$, $\approx$ 140 pores are invaded (see Fig. \ref{fig:poreNumber}c) for efficiency $D_{e} \approx 82.5\%$ (see Fig. \ref{fig:de_df_pd}a); extrapolating $D_{e} = 1$ provides a representative value of $n \approx 170$.
$\lambda$ in the expression for $k_{i}$ is determined empirically to be $\lambda = \overline{D_{e}} [{l_{f} w_{f}}/{l w}] [{w_{f}}/{w}]$. 
%\footnote{RH: (1) add ref to where this comes from--is this the right one? (2) avoid using "frac" inline--it does not show well, I changed here and in many other instances to simple fractions x/y, please change elsewhere. [SP:] this is empirical as we stated and explained the meaning of this expression below.}
The product ${l_{f}w_{f}}$ is the approximate area occupied by a single finger, where $l_{f}$ and $w_{f}$ are the length and width of the finger, respectively. We assume $w_{f} = 1~\textrm{mm}$ and $l_{f} = l_{p} \tau$ where $l_{p}$ is the length of the porous media over which the fingers develop and $\tau$ is the effect of tortuosity \citep{xu2022packing}. 
%\footnote{can't just say we use that value; need to add ref--this is the right one? [SP: done.]}
For VF, as the fingers are of the size of the entire domain length, we use $l_{p} = 14.05~\textrm{mm}$. For VF/CD, as the fingers roughly exist over half the length of the porous medium, we use $l_{p} = 7~\textrm{mm}$.
To account for the effect of tortuosity, we use $l_{f} = 1.5 l_{p}$.  
For VF, $l_{f} \approx 20~\textrm{mm}$ and for VF/CD, $l_{f} \approx 10~\textrm{mm}$. 
We normalize ${l_{f}w_{f}}$ by the area $l w$ (medium dimensions).
%in the equation to determine $\lambda$. 
$\overline{D_{e}} w_{f}/w$ is an empirical parameter used to determine $\lambda$ such that $M$ in Eq. \eqref{eq:stabA} does not become negative. We obtain 
$\overline{D_{e}} \approx 32.5\%$ for VF, $\overline{D_{e}} \approx 65\%$ for VF/CD are the average displacement efficiencies. 
Assuming a proportionality between the effective permeabilities of the two fluids 
$k_{d} = A k_{i}$ \citep{ahmed2018reservoir}, we manipulate Eq. \eqref{eq:stabilityAna} to obtain 
\begin{equation}
    - \frac{\alpha}{\gamma} \bigg[ \frac{1}{A} + M\bigg] + \textbf{U} \bigg[ \frac{1}{A} - M\bigg] = \frac{4 \sigma cos(\theta) \phi \lambda}{\mu_{def} n}.
    \label{EQ_crossover}
\end{equation}
The crossover between flow regimes is expected to occur when $\alpha = 0$ \citep{rabbani2018suppressing}. Substituting $\alpha = 0$ in Eq. \ref{EQ_crossover} provides the following condition
\begin{equation}
    M = \frac{1}{A} - \frac{4 \phi \lambda cos (\theta)}{Ca\cdot n}.
    \label{eq:stabA}
\end{equation}
In the above equation, $A$ is determined empirically. 
For VF, the invading fluid propagates through the porous medium in the form of thin fingers. Therefore, most of the pores remain occupied by the defending fluid. As the effective permeabilities $k_{i}$ are function of phase saturation $S_{i}$ \citep{ahmed2018reservoir}, this imply that the transition from VF to VF/CD occurs at $A = {k_{d} (S_{d})}/{k_{i} (S_{i})} >1$. For this case, we assume $A = \frac{1}{\overline{D_{e}} \phi} \approx 10$. 
While considering VF/CD to CD, more than half of the porous medium is occupied by the invading fluid essentially making $A < 1$. For this case, we assume $A = \overline{D_{e}} \phi \approx 0.2$. 
Substituting all the above variables in Eq. \eqref{eq:stabA} gives the threshold viscosity ratio (the boundary between regimes) at which transition between flow regimes occur indicated by green circles in Fig. \ref{fig:phaseDia}. 
We note that the threshold $M$ increases with $\theta$, in particular for the crossover between VF and VF/CD (Fig. \ref{fig:phaseDia}). 
Interestingly, for the idealized geometry in \citet{primkulov2021wettability} the transition from VF to CD occurred at $M \approx 0.5$ irrespective of the wettability, without an intermediate VF/CD regime which could potentially be due to the relatively simplistic nature of the models used in PNM. 
}

\begin{acknowledgments}
HR acknowledges support from Texas A\&M University at Qatar and Qatar Foundation. RH acknowledges support from the Engineering and Physical Sciences Research Council (EP/V050613/1).  
\end{acknowledgments}

%\hl{[RH: I changed formatting of citations to generic author/year, and fixed few errors in bibtex file.]}

%\bibliographystyle{alpha}
%\bibliographystyle{elsarticle-harv}
\bibliography{bib_interplay_viscosity_wettability_V3.bib}

\end{document}